\documentclass[aps,prb,twocolumn,showpacs,letterpaper,superscriptaddress]{revtex4-1}

\usepackage[colorlinks=true, citecolor=blue, linkcolor=blue, urlcolor=blue]{hyperref}

\usepackage{graphicx,dcolumn,longtable,epsfig}
\usepackage[usenames]{color}
\usepackage{amssymb}
\usepackage{amsmath}
\usepackage{bm}
\usepackage{footnote}
\usepackage{float}
\usepackage{subfigure}
\usepackage{color}

\usepackage{ulem}
\usepackage[T1]{fontenc}

\input epsf.tex

\def\bea{\begin{eqnarray}}
\def\eea{\end{eqnarray}}
\def\be{\begin{equation}}
\def\ee{\end{equation}}

\begin{document}
\title{Peierls/Su-Schrieffer-Heeger polarons in two dimensions}
\author{Chao Zhang}
\affiliation{Department of Physics, University of Massachusetts, Amherst, Massachusetts 01003, USA}
\author{Nikolay V. Prokof'ev}
\affiliation{Department of Physics, University of Massachusetts, Amherst, Massachusetts 01003, USA}
\author{Boris V. Svistunov}
\affiliation{Department of Physics, University of Massachusetts, Amherst, Massachusetts 01003, USA}
\affiliation{National Research ``Center Kurchatov Institute,'' 123182 Moscow, Russia}
\affiliation{Wilczek Quantum Center, School of Physics and Astronomy and T. D. Lee Institute, Shanghai Jiao Tong University, Shanghai 200240, China}

\begin{abstract}
Polarons with different types of electron-phonon coupling have fundamentally different properties.
When the dominant interaction is between the electron density and lattice displacement,
the momentum of the ground state does not change and the polaron gets exponentially heavy at strong coupling.
In contrast, one-dimensional Peierls/Su-Schrieffer-Heeger (PSSH) polarons with
interaction originating from displacement-modulated hopping feature a shift of the
ground-state momentum to finite values and moderate values of effective mass as coupling is increased \cite{Marchand:2010dk}.
Based on Diagrammatic Monte Carlo method, we investigate whether
unusual properties of PSSH polarons depend on the type of the displacement-modulated hopping
and to what degree they survive in higher dimension.
We study two different PSSH models: with bosonic degrees of freedom
residing on sites (model A) and bonds (model B) of the two-dimensional square lattice.
For model A, we find that in both adiabatic and intermediate regimes, the momentum of the ground state experiences a continuous transition from zero to a finite value as a function of coupling strength. The transition is driven by quadratic instability of the dispersion function, implying that  effective mass diverges at the critical point, and then decreases in an anisotropic fashion with increasing coupling.
Unexpectedly, for model B, the momentum of the ground state always stays at zero and
the effective mass increases monotonously with coupling. The increase
is far from exponential and tends to level-off at strong interaction, resulting in relatively light polarons. Having light polarons in the strong coupling regime is crucial for the bi-polaron
mechanism of high-temperature superconductivity \cite{Sous:2018di}.
\end{abstract}

\pacs{}
\maketitle

\section{Introduction}
\label{sec:sec1}

Polarons form a special class of stable quasiparticles emerging as a result of renormalization---often quite dramatic---of bare particle properties by a quantum environment.
Depending on the nature of the particle, environment, and type of coupling,
there are numerous examples of different polarons across all fields of physics: electron-phonon polarons~\cite{Landau33,Frohlich50,Feynman:1955du, Schultz:1959el, Holstein59, Alexandrov:1999fy, Holstein2000},
spin-polarons~\cite{BR70,Nagaev:1974jb, Mott:2006fa}, Fermi-polarons~\cite{Bulgac:2007bl, Lobo:2006dc, Prokofev:2008jz, Prokofev:2008it}, protons in neutron rich matter \cite{protons}, etc.

One of the reasons the electron-phonon polaron problem keeps attracting a lot of attention is the search for
the bi-polaron mechanism of high-temperature superconductivity when $T_c$ is determined by the Bose condensation
of preformed electron pairs. However, reaching large values of $T_c$ for bi-polarons in models with strong
density-displacement coupling is problematic because of exponentially large effective masses\cite{Chakraverty}. To see why, consider the Holstein model \cite{Holstein59}
on the simple cubic lattice when
 \[
 H = H_{e}+H_{\rm ph}+H_{\rm int}\, ,
 \]
\begin{eqnarray}
H_{e} & = & - t       \sum_{<ij>} \, (c_{j}^{\dagger} c_{i}^{\;} +H.c.),
\nonumber \\
H_{\rm ph} &=& \omega_{\rm ph} \sum_{i} \, (b_{i}^{\dagger} b_{i}^{\;} +1/2),
\label{H0} \\
H_{\rm int} &=& g  \sum_{i} \, c_{i}^{\dagger} c_{i}^{\;} X_{i} \,,
\qquad
X_{i}=b_{i}^{\;} + b_{i}^{\dagger}\, ,
\label{Holstein}
\end{eqnarray}
with the frequency of the local phonon mode $\omega_{\rm ph}$ much smaller than the particle
bandwidth $W =12t $. Here $b_{i}^{\;}$ ($c_{i}^{\;}$) are the optical phonon (electron)
annihilation operators on site $i$,  $t$ is the hopping amplitude between the nearest neighbor sites
(we use it as the unit of energy),
and $g$ is the strength of the electron-phonon interaction (EPI) of the density-displacement type.
On the one hand, by treating EPI perturbatively, one obtains light polarons with
slightly renormalized band bottom $E_G \sim -W/2 - c g^2/W$, where $c$ is a
numerical coefficient of the order of unity.
On the other hand, a localized electron gains interaction energy
$E_{\rm loc} = -g^2/2\omega_{\rm ph}$, and the overlap integral between the phonon states
adjusted to different electron positions is given by $o(g) = \exp[ -(g/\omega_{\rm ph})^2 ]$.
At the single polaron level, these considerations imply that at
$g \approx g_1= \sqrt{\omega_{\rm ph} W } \ll W$
the light polaron state is replaced with the heavy one characterized by exponentially
small effective hopping $t_h (g_1) = t e^{-W / \omega_{\rm ph} }$,
or exponentially large effective mass $m^* = 1/2a^2t_h$ where $a$ is the lattice constant
chosen to be the unit of length.
The self-trapping crossover---from light to heavy polaron---is sharp and takes place when the light
polaron state is still in the perturbative regime, making the entire argument quantitatively accurate.

When these considerations are generalized to the tightly bound bi-polaron state that gains
interaction energy $E_{2} = -2g^2/\omega_{\rm ph}$,
one finds that the transition to the heavy bi-polaron state takes place at even weaker coupling,
$g \approx g_2  = \sqrt{W \omega_{\rm ph} /2} < g_1$, but the estimate for the effective hopping
of bi-polarons barely changes,
$t_2 (g_2) = 2 t_h^2 (g_2) /(g_2^2/ \omega_{\rm ph} ) = (1/3) t e^{-W / \omega_{\rm ph} }$
(in this regime, bi-polarons move by first breaking the pair). Once bi-polarons are formed,
their effective mass keeps increasing exponentially with $g^2$.
Since $T_c$ is inverse proportional to $m^*$, the the conventional bi-polaron mechanism is not viable.
Repulsive Coulomb interactions push the value of $g_2$ further upwards.

Remarkably, the situation radically changes when the dominant EPI originates from the displacement-modulated hopping, or Peierls/Su-Schrieffer-Heeger (PSSH) coupling:
\be
H_{\rm int} = g \sum_{<ij>} (c_{j}^{\dagger} c_{i}^{\;} +{\rm H.c.})
( X_{i}^{<ij>} - X_{j}^{<ij>}) \;\; \mbox{(model A)}.
\label{modelA}
\ee
Here $X_{i}^{<ij>} =  b_{i}^{<ij>}+(b_{i}^{<ij>})^{\dagger}$
is the dimensionless displacement of the optical mode vibrating along the $<ij>$ bond;
i.e., we now have $d$ bosonic modes on each site in $d$ dimensions.
In $d=1$, both the polaron and bi-polaron states were found \cite{Marchand:2010dk,Sous:2018di}
to remain relatively light even in the strong coupling regime because electrons can gain
interaction energy only by moving between the lattice sites.
Following existing convention, we define the dimensionless coupling constant as
\be
\lambda \, =\,  {2g^2 \over t\, \omega_{\rm ph}} ,
\label{lambda}
\ee
with strong coupling regime corresponding
to $\lambda \gtrsim 1$.
This  potentially opens the door for the bi-polaron mechanism of
high-temperature superconductivity \cite{Sous:2018di}. PSSH bi-polarons are also supposed to
be less sensitive to local repulsive interactions of the Hubbard type.

However, the results reported in Refs.~\cite{Marchand:2010dk,Sous:2018di} were limited to
the one-dimensional chain, and bi-polarons were studied only in the antiadiabatic regime
$\omega_{\rm ph}=3t \sim W$, when the phonon degrees of freedom should be rather
considered as ``fast''  than ``slow''  with respect to the electron motion.
Thus, before the discussion of (and search for)
the bi-polaron mechanism of high-temperature superconductivity can be projected on
realistic materials, one needs to understand (i) to what extent the intriguing
results for PSSH polarons (including the change of the ground state momentum)
survive in higher dimensions, (ii) whether the picture holds in the most relevant
adiabatic regime $\omega_{\rm ph} \ll W$, and (iii) how sensitive it is to model variations.

Indeed, an alternative way to model the displacement-modulated interaction is
by placing optical phonon degrees of freedom on lattice bonds \cite{modelB}
\be
H_{\rm int} = g \sum_{<ij>} (c_{j}^{\dagger} c_{i}^{\;} +{\rm H.c.}) \,
X_{<ij>} \qquad  \mbox{(model B)},
\label{modelB}
\ee
\[
X_{<ij>} =  b_{<ij>}+b_{<ij>}^{\dagger} .
\]
Despite close similarities, including severe sign-problem in the path-integral representation,
models A and B have different microscopic structure, and thus may radically deviate from each other at strong coupling.

In this work, we employ the Diagrammatic Monte Carlo (DiagMC) method to study ground-state
properties of two-dimensional PSSH polarons in models A and B, in both adiabatic and anti-adiabatic regimes.
The DiagMC technique for polarons is well
established~\cite{Prokofev:1998fj, Mishchenko:2000co, Mishchenko:2019jb, Marchand:2010dk, Mishchenko:2014bn}
and its advantage over the path integral representation for PSSH polarons comes from much better
handling of sign-alternating contributions in momentum space.

Having light polarons in the strong coupling regime is crucial for the bi-polaron
mechanism of high-temperature superconductivity \cite{Sous:2018di}. With this context in mind, the central quantity of our interest
is the effective mass, which we extract from the energy dispersion (obtained from the polaron Green's function).
Our main result is that two-dimensional PSSH polarons, regardless of the model, have relatively
light effective masses at strong coupling even in the adiabatic regime $\omega_{\rm ph}/W \ll 1$.
We did not find evidence for exponential growth of $m^*$ up to the largest coupling constant
we were able to simulate reliably, in sharp contrast with properties of the Holstein polarons.

It turns out that the two PSSH models, despite similarities in the type of EPI, have
radically different properties in the ground state at strong coupling.
In model A, in both adiabatic and intermediate regimes, the momentum of the ground state experiences a continuous transition from zero to a finite value as a function of coupling strength. The transition is driven by quadratic instability of the dispersion function, implying that  effective mass diverges at the critical point, and then decreases in an anisotropic fashion with increasing coupling.
An alternative scenario of a transition to a finite-momentum ground state is the scenario of competing sectors, when the energy at a certain finite momentum drops below the energy of the zero-momentum state. Our data does not support the competing sectors scenario.

Unexpectedly, for model B, the momentum of the ground state always stays at zero and
the effective mass increases monotonously with coupling. The increase
is far from exponential and tends to level-off at strong interaction, resulting in relatively light polarons.

The rest of the paper is organized as follows.
In Sec.~\ref{sec:sec2}, we reformulate our models in momentum representation and describe the
configuration space of Feynman diagrams simulated by the DiagMC method.
In Sec.~\ref{sec:sec3}, we introduce the protocol of the Green's function data analysis that allows us to
extract polaron energies and $Z$-factors for various momenta.
In Sec.~\ref{sec:sec4}, we render the theory of anisotropic effective mass with emphasis on the
case of quadratic instability in the $D_{4h}$-symmetric system and the corresponding fitting ansatzes. In Sec.~\ref{sec:sec5},
we present results for the ground-state properties and discuss how they change with the model
and adiabatic regime. We conclude and discuss perspectives in Sec.~\ref{sec:sec6}.

\section{Diagrammatic Monte Carlo setup}
\label{sec:sec2}

%
\begin{figure}
\includegraphics[ width=0.48\textwidth]{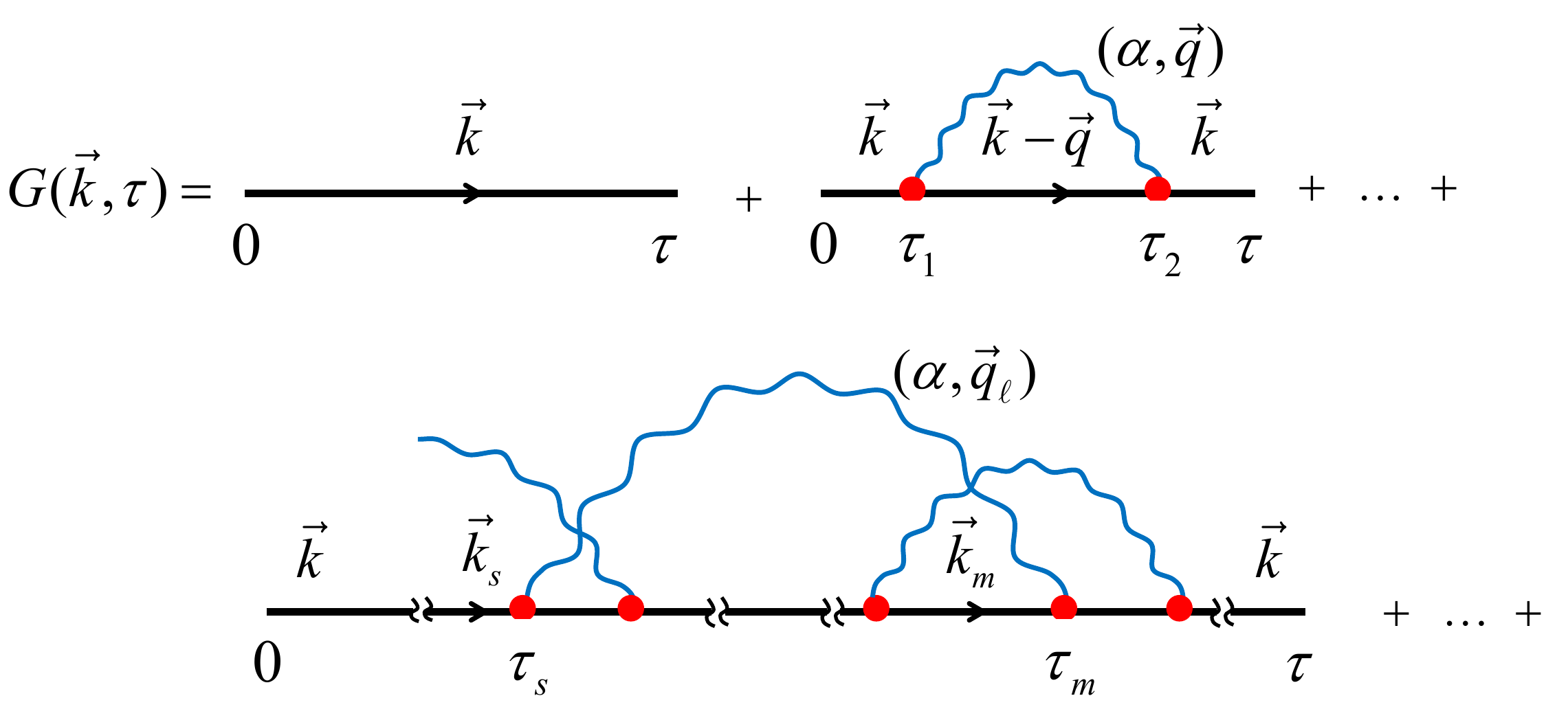}
\caption{(color online) Diagrammatic expansion for the polaron Green's function. Straight (wavy)
lines represent bare particle (phonon) propagators $G_0$ ($D_0$), and dots stand for interaction vertexes
(see text).
}
\label{FIG0}
\end{figure}

In momentum representation, the non-interacting system is characterized by the tight-biding dispersion relation
\[
 \epsilon_{\mathbf{k}}=-2 t [\cos (k_x a)+ \cos (k_y a)] \]
with the bandwidth, $W=8t$, and effective mass at zero momentum, $m_0=1/2ta^2$, for the particle,
and two dispersionless optical modes $\omega_{\alpha, \mathbf{q}} =\omega_{\rm ph}$ for lattice vibrations.
The corresponding adiabatic parameter is then defined by $\gamma= \omega_{\rm ph}/8t$.
We consider $t$ as the unit of energy.

The interaction term for both PSSH models can be written as
\be
H_{\rm int} = V^{-1/2} \sum_{\mathbf{k}, \mathbf{q}, \alpha} \left[
M_{\alpha}(\mathbf{k}, \mathbf{q}) \,
c_{\mathbf{k}- \mathbf{q}}^{\dagger} c_{\mathbf{k}}^{\;} \, b_{\alpha,\mathbf{q}}^{\dagger} + {\rm H.c.} \right] .
\label{modelsAB}
\ee
Here $V$ is the number of lattice sites, $\alpha = 1,2$ labels vibrational modes responsible for
modulation of the hopping amplitude along bonds in directions $\hat{x}$ and $\hat{y}$, respectively.
The key difference between the PSSH and Holstein models
is that in the former the interaction amplitude, $M_{\alpha}(\mathbf{k}, \mathbf{q})$, depends explicitly on the
incoming electron momentum. In Holstein and Fr\"{o}hlich models this dependence is absent, leading to the
sign-free diagrammatic expansion because the product of amplitudes corresponding to the creation
and annihilation of the phonon excitation is trivially positive: $M_{\alpha}(\mathbf{q}) \, M_{\alpha}^*(\mathbf{q})= |M_{\alpha}(\mathbf{q})|^2$.
This is no longer the case for PSSH models.  Here the product
$M_{\alpha}(\mathbf{k}_s, \mathbf{q}) \, M_{\alpha}^*(\mathbf{k}_m+\mathbf{q},\mathbf{q})$ is sign-alternating
(for higher-order diagrams; see Fig.~\ref{FIG0}) as is easily seen from explicit expressions
\bea
M_{\alpha}(\mathbf{k}, \mathbf{q}) &=& 2ig [ \sin (k_\alpha - q_\alpha) - \sin (k_\alpha) ]
\;\;\mbox{(model A)} , \;\;
\label{modelsAA} \\
M_{\alpha}(\mathbf{k}, \mathbf{q}) &=& 2g \cos (k_\alpha - q_\alpha/2 )
\;\;\mbox{(model B)}  .
\label{modelsBB}
\eea
As a result, the Monte Carlo simulation of the diagrammatic expansion in the momentum representation
suffers from the sign problem, which, however, is not as severe as in the path-integral representation because
the product of vertex functions groups together a number of sign-alternating contributions
(16 for model A and 4 for model B). The other advantage is that size effects are absent altogether.

The diagrammatic expansion for the particle Green's function is illustrated in Fig.~\ref{FIG0}.
Each contribution is a product of functions associated with the graph elements: straight lines
represent bare particle Green's functions,
\[
G_0 (\mathbf{k}_s, \tau_s - \tau_{s-1} ) = \exp \{ - [\epsilon(\mathbf{k}_s)-\mu ](\tau_s - \tau_{s-1} ) \} ,
\]
wavy lines represent bare phonon propagators,
\[
D_0 (\alpha , \mathbf{q}_{\ell}, \tau_m - \tau_s ) = \exp [ -\omega_{\rm ph} (\tau_m - \tau_s ) ],
\]
and dots stand for the interaction vertexes: amplitudes $M_{\alpha}(\mathbf{k}, \mathbf{q})$, or their
complex conjugates. The configuration space sampled by the DiagMC method
includes the polaron momentum $\mathbf{k}$,
the graph duration in imaginary time $\tau$,
the diagram order $n$ (number of phonon lines),
indexes $\alpha_1, \dots , \alpha_n$ and momenta $\mathbf{q}_1, \dots , \mathbf{q}_n$
of the phonon lines (particle momenta are then fixed by the conservation laws),
and the set of imaginary time points $\tau_1, \dots , \tau_{2n}$ for interaction vertexes.
The rest of the technique---except for data processing that needs to be modified for
the sign-alternating expansion and is described next---is standard and closely follows detailed
descriptions provided in Refs.~\cite{Prokofev:1998fj, Mishchenko:2000co}.

\section{Green's function data analysis}
\label{sec:sec3}

\begin{figure}
\includegraphics[ trim=0.4cm 6cm 0cm 7cm, clip=true, width=0.5\textwidth]{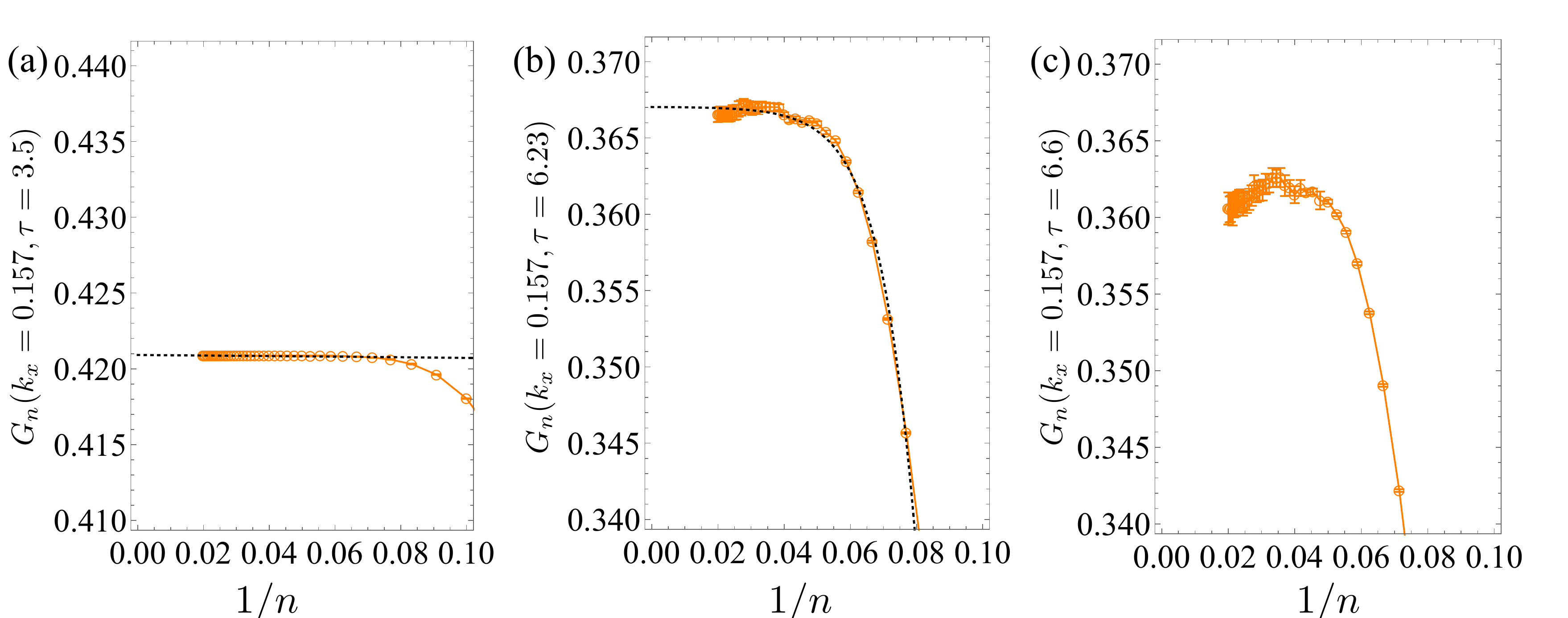}
\caption{(color online)
Green's function dependence on the inverse expansion order for model A
in the adiabatic regime $\gamma=\omega_{\rm ph}/W=1/16$ with $\lambda =2.074$, $ \mu =-4.88$, and $\mathbf{k}=(0.157,0)$.
Simulation data in panels (a), (b), and (c) show how extrapolation of the infinite diagram-order limit
is done for $\tau=3.5$ (a) and $\tau = 6.23$ (b), but becomes problematic at longer times,
$\tau = 6.6$ (c), due to large sign-related statistical errors.
The black dotted line is the fit to the logistic function.
}
\label{FIG9}
\end{figure}

\begin{figure}
\includegraphics[ trim=0.4cm 4.5cm 3cm 5cm, clip=true, width=0.5\textwidth]{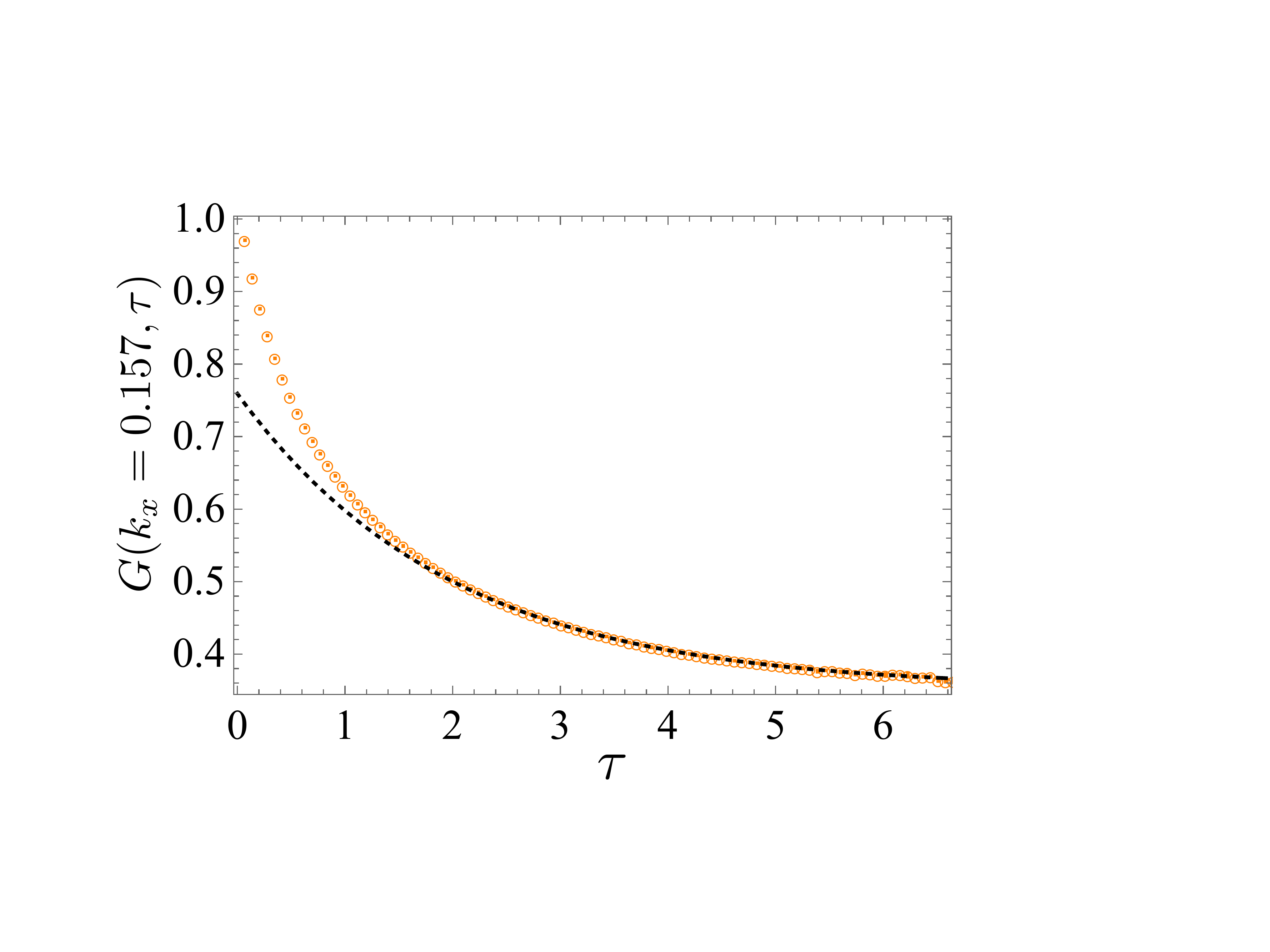}
\caption{(color online)
Extrapolated Green's function dependence on imaginary time for model A in the adiabatic
regime $\gamma=1/16$ when $\lambda=2.074$, $\mu=-4.88$, and $\mathbf{k}=(0.157,0)$.
The black dotted line is a fit to the exponential dependence with leading correction:
$f(\tau ) = 0.34513 e^{0.00261 \tau} [1+1.19966 e^{-0.5 \tau}]$, see Eq.~(\ref{Gfit}).
Each point is based on the infinite diagram-order extrapolation shown in Fig.~\ref{FIG9}.
Error bars are shown and are smaller than the symbol size.
}
\label{FIG8}
\end{figure}

The diagrammatic expansion for $G(\mathbf{k}, \tau )$ for lattice polarons converges for any values of
momentum and imaginary time because the factorial number of different diagram topologies, $(2n-1)!!$, is
well overcompensated by the integration measure of time-ordered interaction vertexes, $\propto \tau^{2n}/(2n)!$
(after momentum integration all functions remain non-singular on the $\tau$-axis).
For sign-positive expansions this observation implies that highly accurate data for $G(\tau )$
can be obtained for long values of $\tau $ when projection to the ground state properties is perfect
for all practical purposes. The average expansion order for $G(\tau)$ does increase linearly with $\tau$,
but the computational cost of sampling the corresponding contributions with small error-bars is very mild
due to self-averaging effects (multiple repeated instances of the proper self-energy insertions).

The situation radically changes for sign-alternating series because now Monte Carlo sampling comes with much larger
error bars that grow exponentially with the diagram order. Thus, for a given simulation
time, precise data can be obtained only up to some limited diagram order
(in practice it is $n \lesssim 50$ in the strong coupling regime),
and as a consequence, only up to some limited imaginary time $\tau_{\rm max}$.
This situation is illustrated in Fig.~\ref{FIG9} for model A in the adiabatic regime $\gamma = 1/16$.
The coupling strength was chosen to be slightly larger than the critical value for
transition to the ground state with finite momentum (see Sec.~\ref{sec:sec41}), $\lambda=2.074 > \lambda_c \approx 2.01$.
For relatively short imaginary time $\tau = 3.5$  (still larger than $\omega_{\rm ph}^{-1}$),
the exponential convergence is evident, and
extrapolation to the infinite diagram order limit by fitting the data for $n\ge 13$ to the logistic function,
\[
y(x)= \frac{a}{1+b e^{-c /x}}  \qquad \qquad (x=1/n) ,
\]
results in an accurate answer $G(\mathbf{k},\tau=3.5)=0.4209(5)$ for $\mathbf{k}=(0.157,0)$.
For the same parameter set at $\tau=6.23$, the Green's function convergence is achieved with visibly larger
statistical error bars to which one has to add a comparable extrapolation error,
$G(\mathbf{k},\tau=6.23)=0.367(2)$; see Fig.~\ref{FIG9}(b).
Finally, at $\tau=6.6$, see Fig.~\ref{FIG9}(c), the statistical errors become too large before the
convergence is reached, at which point we have established the largest simulation time suitable for further
analysis ($\tau_{\rm max}=6.5$ for the parameter set discussed).

To extract the polaron energy, $E(\mathbf{k})$,  and $Z(\mathbf{k})$-factor at momentum $\mathbf{k}$
from the Green's function dependence on imaginary time, see Fig.~\ref{FIG8}, we perform the following
analysis. In the asymptotic limit $\tau \to \infty$, this dependence is governed by the ground state
in the corresponding momentum sector, as follows from the spectral Lehman representation.
For the stable (non-decaying) quasiparticle state, we have
\begin{equation}
G(\mathbf{k},\tau \to \infty) \rightarrow Z(\mathbf{k}) e^{-[E(\mathbf{k})-\mu] \tau} \, .
\label{limit}
\end{equation}

In the absence of additional stable quasiparticle states,
the spectral density is zero up to the threshold, $E_{\rm th} = E(\mathbf{k})+ \omega_{\rm ph} $, for
emission of the optical phonon. Thus, the leading finite-$\tau$ correction to Eq.~(\ref{limit})
starts with and additional exponential factor $e^{-\omega_{\rm ph} \tau}$. Since our data for sign-alternating
expansions cannot be extended to arbitrary long imaginary times, contrary to the
situation for Fr\"{o}hlich and Holstein polarons, the corresponding correction is included in fitting the data
at large enough times:
\begin{equation}
G(\mathbf{k},\tau \gg \omega_{\rm ph}^{-1}) \, \rightarrow \, Z(\mathbf{k}) e^{-[E(\mathbf{k})-\mu ] \tau}  [1 + C e^{-\omega_{\rm ph} \tau} ] \, .
\label{Gfit}
\end{equation}
A typical example is presented in Fig.~\ref{FIG8}. The quality of the fit (dotted line)
ensures that there are no additional stable states with measurable $Z$-factors at energies $E < E_{\rm th}$.
The final result for $\mathbf{k}=(0.157,0)$ extracted from this set of data is
$E(\mathbf{k})=-4.883(2)$, and $Z(\mathbf{k})=0.345(3)$.

Our attempts to reduce the severity of the sign-related problem by grouping diagrams,
sampling the proper self-energy instead of the Green's function
and employing the skeleton formulation with self-consistent feedback in the form of the Dyson equation,
produced data of the same quality as sampling the bare Greeen's function expansion.

\section{Principal Effective Masses}
\label{sec:sec4}

\subsection{General relations}

At the point of extremum (a minimum, a maximum, or a saddle point) ${\bf k}={\bf k}_0$, the energy $E({\bf k})$ can be expanded as
\be
E({\bf k})\, =\, E({\bf k}_0)\,  +\,  {1\over 2}\sum_{ij} Q_{ij} \xi_i \xi_j \, +\, o(\xi^2)\, ,
\label{BS1}
\ee
where $\vec{\xi}= {\bf k}-{\bf k}_0$ (in components: $\xi_i = k_i-k_{0i}$) and
\be
Q_{ij} \, =\, Q_{ji} \, =\,\left.  {\partial^2 E\over \partial k_i \partial k_j} \right\vert_{{\bf k}={\bf k}_0} \, .
\label{BS2}
\ee
The inverse principal values of the real symmetric tensor $Q_{ij}$ are called principal effective masses, $m_*^{(\nu)}$, $\nu=1,2,\ldots, d$ . The corresponding unit
eigenvectors $\hat{n}^{(\nu)}$ define (the directions of) the principal axes of the tensor $Q_{ij}$, implying the following representation
\be
Q_{ij} \, =\, \sum_{\nu=1}^d \, {n^{(\nu)}_i n^{(\nu)}_j \over m_*^{(\nu)}} \, .
\label{BS3}
\ee
With this parameterization, Eq.~(\ref{BS1}) becomes
\be
E({\bf k})\, =\, E({\bf k}_0)\,  +\,  \sum_{\nu=1}^d \, {\left[ \vec{\xi} \cdot \hat{n}^{(\nu)} \right]^2 \over 2 m_*^{(\nu)}}\, +\, o(\xi^2)\, .
\label{BS4}
\ee

In a typical situation like ours, the principal axes, $\hat{n}^{(\nu)}$, are known {\it a priori} by the symmetry of the problem, and the principal effective masses are readily found by one of the two simple procedures based on Eq.~(\ref{BS4}). The first procedure is
a direct numeric evaluation of $1/ m_*^{(\nu)}$ from the second-order partial derivative of  $E({\bf k})$ along the principal axis $n^{(\nu)}$ taken at point ${\bf k}_0$, which is also found numerically.
The second procedure is fitting the data for $E({\bf k})$ to the ansatz (\ref{BS4}) with ${\bf k}_0$, $E({\bf k}_0)$, and $m_*^{(\nu)}$ treated
as free fitting parameters.

On approach to the point of quadratic instability, the tensor $Q_{ij}$ vanishes and ansatz (\ref{BS4}) becomes progressively poor.
The procedure of finding ${\bf k}_0$ also becomes problematic in view of the dispersion $E({\bf k})$ flattening
at small momenta. In this situation, we fit $E({\bf k})$ with a more complex ansatz properly capturing the quadratic instability,
compute $Q_{ij}$ analytically, and obtain the principal effective masses from the formula:
\be
{1\over m_*^{(\nu)}} \, =\, \sum_{ij} n^{(\nu)}_i \, Q_{ij} \, n^{(\nu)}_j \, .
\label{BS5}
\ee

\subsection{Implications of $D_{4h}$ symmetry}

Consistent with the $D_{4h}$ symmetry of the problem, we observe numerically that dispersion minima always satisfy the condition
\be
|k_{0y}| \, =\, |k_{0x}| .
\label{BS6}
\ee
Reflection about the axis $k_{y} \, =\, k_{x}$ (or $k_{y} \, =\,- k_{x}$) preserves the position of the point ${\bf k}_0$.
The $D_{4h}$ symmetry then requires that the principal axes be preserved as well, implying that
one can always choose them as
\be
\hat{n}^{(1)} \, =\, \left({1\over \sqrt{2}},\, {1\over \sqrt{2}}\right) \, , \qquad \hat{n}^{(2)} \, =\,  \left({1\over \sqrt{2}},\,- {1\over \sqrt{2}}\right) \, ,
\label{BS7}
\ee
for non-negative components of ${\bf k}_0$. When the off-diagonal element $Q_{xy}$ is zero, the
spectrum is degenerate and Eq.~{\ref{BS7}} remains one of the valid choices.
[Note that Eqs.~(\ref{BS6}) and (\ref{BS7}) hold true also for ${\bf k}_0=0$.]
In accordance with (\ref{BS5}) we then have
\be
{1\over m_*^{(1)}} \, =\, Q_{xx} + Q_{xy}  \, , \qquad {1\over m_*^{(2)}} \, =\, Q_{xx} - Q_{xy} \, .
\label{BS8}
\ee
We took into account that $Q_{xx} = Q_{yy}$ by the $D_{4h}$ symmetry.

At  ${\bf k}_0=0$ the $D_{4h}$ symmetry enforces
\be
Q_{xy}\, =\, 0  \qquad \qquad  (\mbox{at}~~{\bf k}_0=0) \, ,
\label{BS9}
\ee
leading to the isotropic effective mass:
\be
{1\over m_*^{(1)}} \, =\, {1\over m_*^{(2)}} \, =\, Q_{xx}   \qquad \qquad  (\mbox{at}~~{\bf k}_0=0).
\label{BS10}
\ee

\subsection{Quadratic instability in the $D_{4h}$-symmetric case}

The quadratic instability of the ${\bf k}_0=0$ energy minimum in the $D_{4h}$-symmetric system is captured by the following polynomial form
\be
E({\bf k})\, =\, E_0 + A (k_x^2 + k_y^2) + B (k_x^4 + k_y^4) + C k_x^2  k_y^2 \, . \quad
\label{BS11}
\ee
The critical point is the point where the coefficient $A$ nullifies, changing its sign from positive
(stable minimum at ${\bf k}_0=0$) to negative (maximum at ${\bf k}_0=0$). On approach to the  critical point, the description of  transition---evolution of the energy minima and effective masses---in terms of  Eq.~(\ref{BS11}) becomes asymptotically exact
because it is nothing but the Taylor expansion of $E({\bf k})$ in powers of $k_x$ and $k_y$
up to all the leading/relevant terms. An important assumption (verified numerically) is that the
quartic part is stable. Rewriting the quartic part as
\[
B (k_x^4 + k_y^4) + C k_x^2  k_y^2\, =\, B (k_x^2 - k_y^2)^2 + (C+2B)k_x^2  k_y^2\, ,
\]
we see that the necessary and sufficient condition for the the quartic form to be stable is:
\be
B > 0 \qquad \mbox{and} \qquad C > -2B\, .
\label{BS12}
\ee
By rewriting the quartic part as
\[
B (k_x^4 + k_y^4) + C k_x^2  k_y^2\, =\, B (k_x^2 + k_y^2)^2 + (C-2B) k_x^2  k_y^2\, ,
\]
we see that the sign of $(C-2B)$ controls the positions of the energy minima. For $C < 2B$
the minima are along the diagonal directions
\be
|k_{0x}|\, =\, |k_{0y}| \qquad \qquad (C < 2B).
\label{BS14}
\ee
Otherwise they are along the $\hat{x}$ and $\hat{y}$ axes
\be
|k_{0x}| \cdot |k_{0y}| \, =\, 0 \qquad \qquad (C > 2B).
\label{BS15}
\ee
At $C=2B$ we would need to take into account higher-order terms in the Taylor expansion.
Our numerics is consistent with the case (\ref{BS14}).

Solving for the minima (\ref{BS14}) and then using (\ref{BS8}) to calculate the principal masses, we get
\be
{\bf k}_0=0\, , \qquad \quad {1\over m_*^{(1)}} = {1\over m_*^{(2)}}  = 2A \qquad \quad (A \geq 0),
\label{BS16}
\ee
\be
k_{0x}^2 = k_{0y}^2= {|A| \over 2B+C} \quad \quad (A \leq 0)  ,
\ee
\be
{1\over m_*^{(1)}}  =  4|A| \, , \quad \quad
{m_*^{(1)} \over m_*^{(2)}} \, =\, {2B-C\over 2B+C} \quad \quad (A \leq 0).
\label{BS17}
\ee

\subsection{Trigonometric ansatz}

In terms of asymptotically exact semi-analytic description of the transition, an interesting alternative to the polynomial
ansatz (\ref{BS11}) is the trigonometric ansatz
\begin{eqnarray}
E({\bf k})\, =\, a + b (\cos k_x + \cos k_y) \qquad   \qquad  \qquad \nonumber \\
+\,  c (\cos 2 k_x + \cos 2k_y) + d \cos k_x  \cos k_y \, ,\qquad
\label{BS18}
\end{eqnarray}
where the coefficients $a,b,c,d$ are obtained by fitting expression (\ref{BS18}) to the numeric data for $E({\bf k})$ at appropriately small values of $k$ in the vicinity of the transition. Qualitatively, the forms (\ref{BS11}) and (\ref{BS18}) are equivalent, since they have exactly the same---minimum necessary---number of independent parameters. At the quantitative level, the ansatz (\ref{BS18}) may work better,
because, as opposed to  (\ref{BS11}), it features proper periodicity in the reciprocal space, meaning that on departure from
the region of small $k$'s its higher-order in $k$ terms may better capture the actual dispersion relation.

Here we present the expressions for the points of minima and the principal effective masses in terms of the coefficients $a,b,c,d$.
As before, we confine ourselves to the relevant to our simulations case when the points of energy minima obey Eq.~(\ref{BS6}) and the relations (\ref{BS7})--(\ref{BS10}) apply.

We have
\be
{\bf k}_0 \, =\, 0 \qquad \mbox{at} \qquad 4c+b+d \leq 0 ,
\ee
\be
{1\over m_*^{(1)}} \, =\, {1\over m_*^{(2)}} \, =\, |4c+b+d | \qquad \quad  ({\bf k}_0 \, =\, 0) \, ,
\label{BS19}
\ee
\be
\cos k_{0\, x} = -{b\over (4c+d)} \quad  \mbox{at} \quad 4c+b+d \geq 0 \, ,
\label{BS20}
\ee
\be
{1\over m_*^{(1)}} \, =\, {(4c+d)^2 - b^2 \over 4c+d} \qquad \quad ({\bf k}_0 \, \neq\, 0) ,
\label{BS21}
\ee
\be
{m_*^{(1)}\over m_*^{(2)}} \, =\,  {4c-d \over 4c+d} \qquad \qquad ({\bf k}_0 \, \neq\, 0) .
\label{BS22}
\ee
The critical point corresponds to
\be
4c+b+d\, =\, 0\, .
\label{BS23}
\ee

\section{Results and discussion}
\label{sec:sec5}

\subsection{Model A: Vibrational modes residing on lattice sites}
\label{sec:sec41}

\begin{figure}
\includegraphics[ trim=15.0cm 0cm 1.5cm 0cm, clip=true, width=0.6\textwidth]{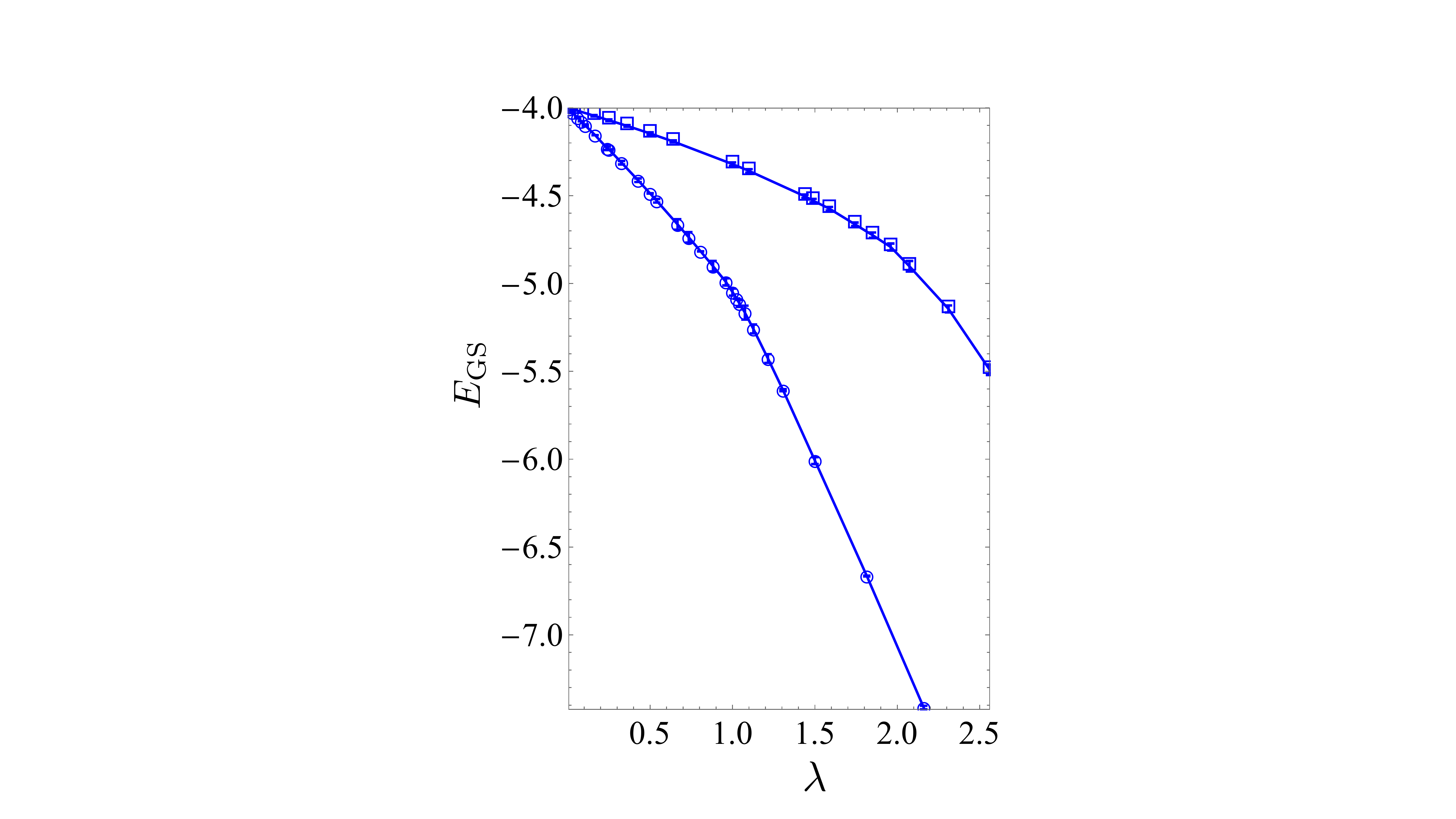}
\caption{(color online)
Ground state energy as a function of coupling strength $\lambda$ for model A in the
adiabatic, $\gamma=1/16$ (squares, upper curve), and intermediate $\gamma=3/8$ (circles, lower curve),
regimes.
}
\label{FIG4}
\end{figure}

In model A, particle hopping is modulated by the relative displacement of atoms located at lattice sites.
Previous work \cite{Marchand:2010dk} found that in 1D the ground state is located at zero momentum only
when the coupling is weak enough. Above the critical value $\lambda_c$, the ground state shifts to finite
values of $\mathbf{k}$. As the coupling constant is increased further, the quasiparticle residue
quickly decays to zero, but
the effective mass goes through a maximum (divergence) and decreases back to relatively small values.
Our simulations confirm that at the qualitative level this picture holds in 2D,
and light polarons exist at strong coupling in 2D as well with one important distinction:
in this regime the effective mass becomes anisotropic.

In Fig.~\ref{FIG4}, we show the ground state energy as a function of $\lambda$ for two values of the adiabatic
ratio  $\gamma = \omega_{\rm ph}/W$. The upper and lower curves correspond to
$\gamma=1/16$ and $\gamma=3/8$, respectively.
As we enter the strong coupling regime, both curves suggest that $E_{\rm GS}(\lambda)$ has a kink at
some critical value: $\lambda_c \approx 2$ for adiabatic and $\lambda_c \approx 1$ for intermediate regimes.
This behavior is typical for ``first-order'' transitions but---as we argue below---in the present context, it is more accurately
described by the quadratic instability of the energy dispersion, implying a continuous transition.

\begin{figure}
\includegraphics [trim=10.0cm 0cm 1.5cm 0cm, clip=true, width=0.6\textwidth]{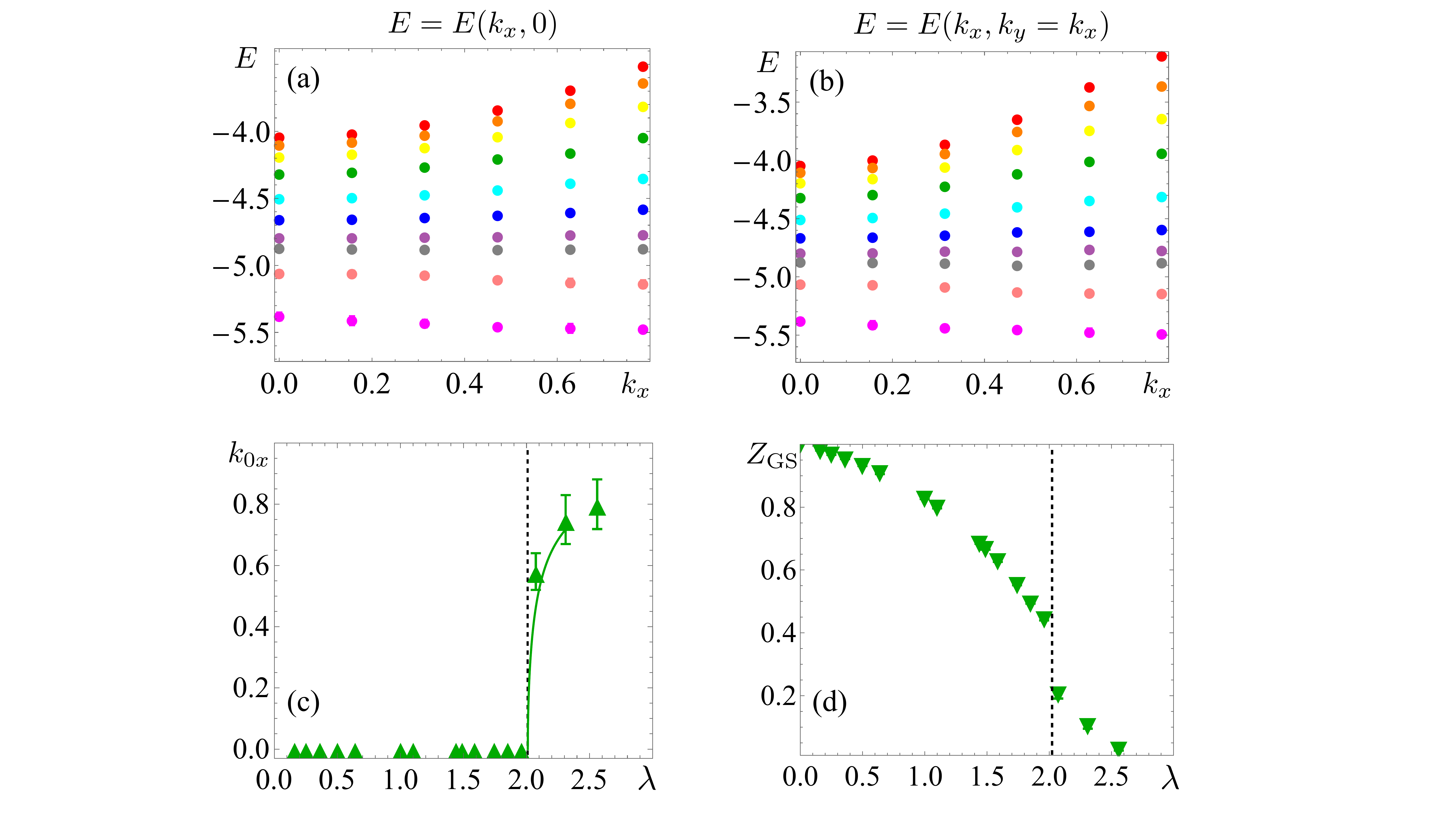}
\caption{(color online)
Polaron properties in the adiabatic regime $\gamma=1/16$ for model A at different couplings. (a)--(b) Energy dispersion up to the decay threshold. Top to bottom: $\lambda=0.16$, 0.36, 0.64, 1.0, 1.44, 1.742, 1.96, 2.074, 2.31, and 2.56. (c) Ground state momentum $k_{0x}$. The semi-analytic solid line is produced by jointly fitting the dispersion functions as explained in the text. (d) $Z_{\rm GS}$-factor.  The vertical dashed line indicates the critical coupling $\lambda_c =2.01(1)$.
}
\label{FIG5}
\end{figure}
\begin{figure}
\includegraphics [trim=10.5cm 0cm 1.5cm 0cm, clip=true, width=0.6\textwidth]{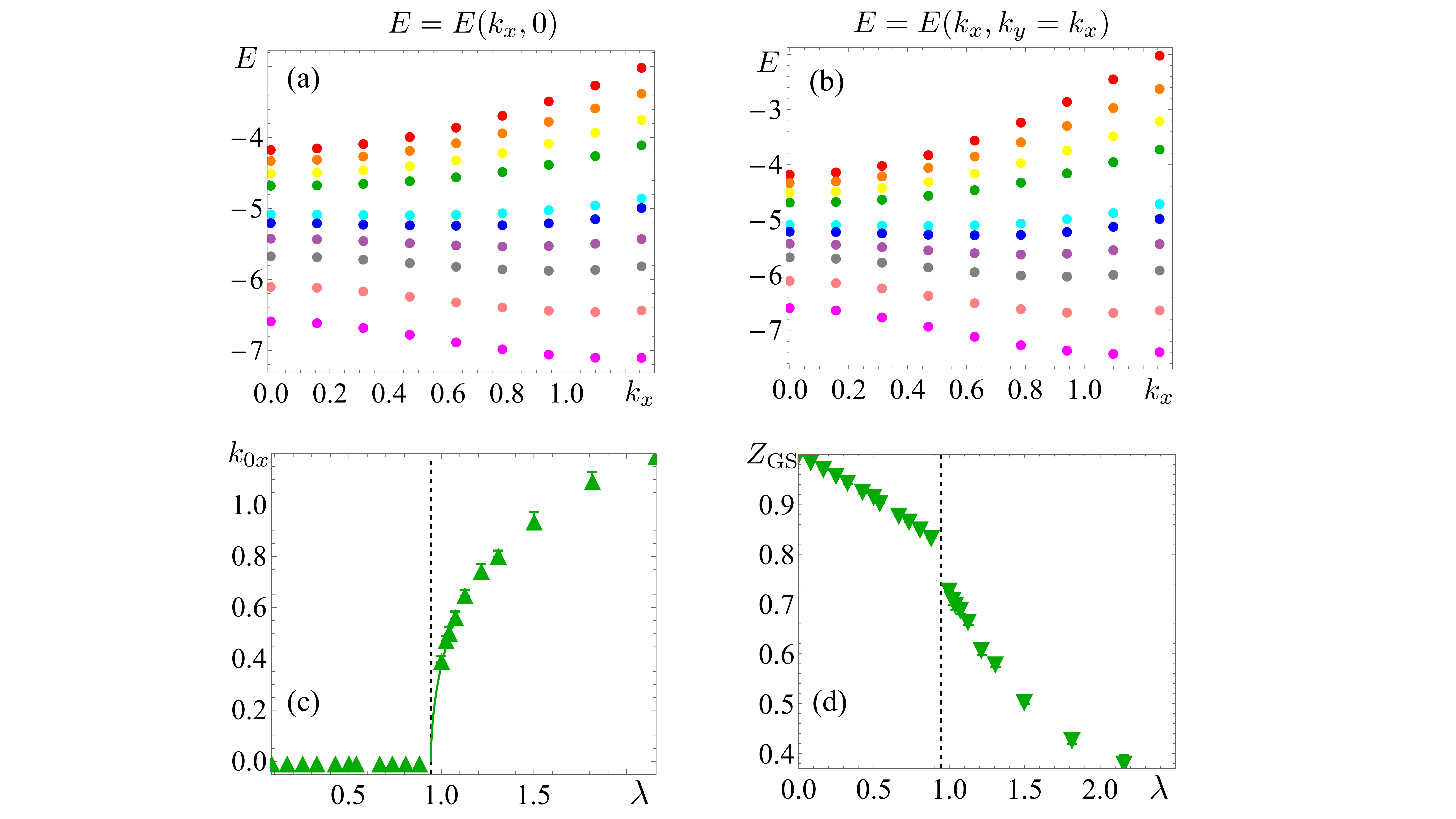}
\caption{(color online)
Polaron properties in the intermediate regime $\gamma=3/8$ for model A at different couplings. (a)--(b) Energy dispersion up to the decay threshold. Top to bottom: $\lambda=0.167$, 0.327, 0.5, 0.667, 1.025, 1.127, 1.307, 1.5, 1.815, and 2.16. (c) Ground state momentum $k_{0x}$.  The semi-analytic solid line is produced by jointly fitting the dispersion functions as explained in the text. (d) $Z_{\rm GS}$-factor. The vertical dashed line indicates the critical coupling $\lambda_c = 0.94(2)$.
}
\label{FIG6}
\end{figure}

The panels (a)--(b) in Figs.~\ref{FIG5} and \ref{FIG6} display the polaron energy dispersion, $E({\bf k})$---for
two characteristic directions in the momentum space---at various coupling parameters. At weak coupling, the minimum
at $\mathbf{k}=0$ is unique. On approach to the critical value $\lambda=\lambda_c$, the function $E({\bf k})$ flattens out,
leading to
heavy polaron states, and then develops a minimum at a finite momentum ${\bf k}={\bf k}_0$, which lies
on the diagonal $k_y=k_x$.  The magnitude of the new ground state momentum ${\bf k}_0$ increases with coupling,
and the quasiparticle residue quickly drops to very small values; see panels (c)--(d) in Figs.~\ref{FIG5} and \ref{FIG6}.
Due to momentum conservation, the different momentum states cannot be mixed, and the transition is sharp, even if
the dispersion relation changes continuously.

With our numeric resolution, it is hard to unquestionably distinguish---by the brute force---between a continuous and a weak discontinuous transitions. Both scenarios are allowed because at small momenta the dispersion relation can be expanded only in even powers
of $k_x$ and $k_y$, as dictated by lattice symmetries. Since we do not find evidence for a
metastable minimum of $E({\bf k})$ emerging at $\lambda < \lambda_c$, the transition
most likely goes through the continuous scenario when the quadratic form becomes non-positive with stabilization provided by  quartic terms.

\begin{figure}
\includegraphics[ trim=15cm 0cm 5cm 0.cm, clip=true, width=0.65\textwidth]{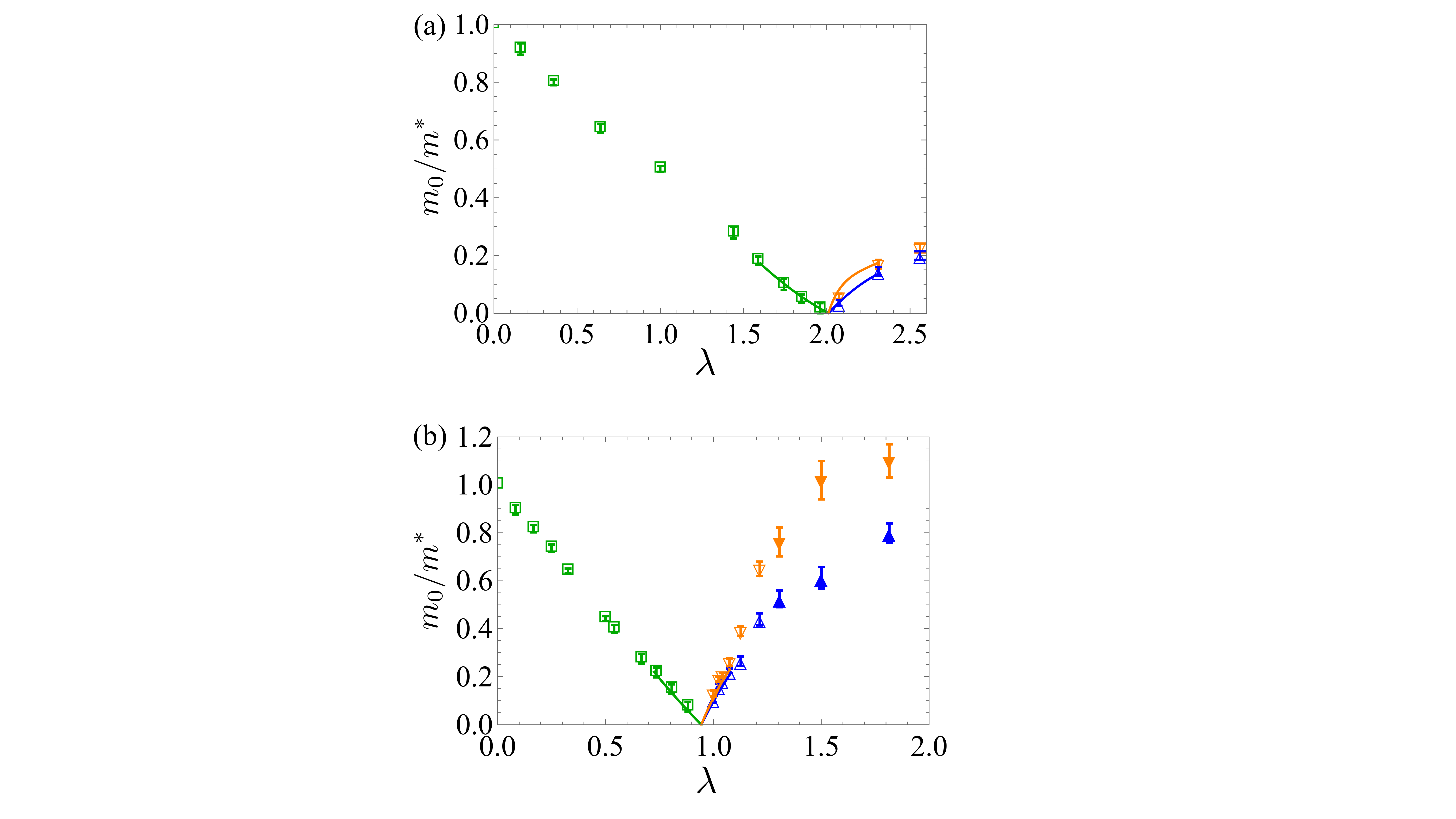}
\caption{(color online)
Principal effective masses as functions of coupling strength $\lambda$ for model A in the adiabatic, $\gamma=1/16$ (a), and intermediate, $\gamma=3/8$ (b), regimes. Open symbols are used for the data points obtained by fitting with the ansatz (\ref{BS18}). The data shown with closed symbols are extracted by fitting with the generic finite-${\bf k}_0$ ansatz (\ref{BS4}), with the principal axes (\ref{BS7}).
The blue up-triangles stand for $1/m_{*}^{(1)}$, the inverse principal effective mass along the diagonal;  the orange down-triangles represent $1/m_{*}^{(2)}$, the inverse principal effective mass perpendicular to the diagonal.  The semi-analytic solid lines are produced by jointly fitting the dispersion functions as explained in the text.
}
\label{FIG7}
\end{figure}

The crucial piece of evidence strongly supporting this scenario is provided by successfully fitting numeric data for $E({\bf k})$ in the vicinity of the critical point by the polynomial and trigonometric ansatzes,
Eqs.~(\ref{BS11}) and (\ref{BS18}), describing the transition driven by quadratic instability (in a $D_{4h}$-symmetric
system). In the vicinity of the critical point, we used the most conservative fitting protocol requiring that the free parameters in Eqs.~(\ref{BS11}) and (\ref{BS18}) are smooth structureless functions of $\lambda$ across the transition point. Specifically, we employed parabolic (and even linear in some cases) ansatzes for these functions with the
coefficients of corresponding polynomials being extracted from jointly fitting dispersion relations
$E({\bf k})$ for a set of $\lambda$'s in the vicinity of $\lambda_c$. We found all our data consistent
with such fitting. Along with strongly supporting the continuous scenario, our protocol naturally produces
semi-analytic results for the evolution of ${\bf k}_0$ and principal masses across the critical point; see solid
lines in panels (c) in Figs.~\ref{FIG5} and \ref{FIG6} and in Fig.~\ref{FIG7}. In particular, note that semi-analytic
curves for $k_{0x}$ are perfectly consistent with the evolution of $k_{0x}$ at $\lambda > \lambda_c$ found from
the energy minimuma at a given value of $\lambda$.

In the adiabatic regime, $\gamma=1/16$, we used ansatz (\ref{BS11}). By the above-described joint fitting protocol,
the coefficients $E_0$, $A$, $B$, and $C$ were found to obey
\begin{eqnarray}
 \gamma=1/16: \qquad \qquad \qquad \qquad \qquad \qquad \qquad \qquad  \qquad \qquad  \nonumber \\
E_0 =-0.42(3)\lambda^2+0.86(4) \lambda-4.89(10), \qquad \qquad \label{semi_1_1}\\
A=[2.01(1)-\lambda][0.86(4)-0.27(2)\lambda], \qquad \qquad \label{semi_1_2}\\
B=0.03(1)\lambda+0.01(1),  \qquad \qquad \label{semi_1_3}\\
C=-0.40(5)\lambda^2+1.91(5)\lambda-2.28(2) . \qquad \qquad \label{semi_1_4}
\end{eqnarray}
In the intermediate regime $\gamma=3/8$, we also employed  the trigonometric ansatz (\ref{BS18}).
The joint fitting protocol resulted in
\begin{eqnarray}
\gamma=3/8: \qquad \qquad \qquad \qquad \qquad \qquad \qquad \qquad  \qquad  \nonumber \\
a =-7.6(2) \lambda^2 + 11.6(3) \lambda -7.0(2),  \qquad \qquad \label{semi_2_1}\\
b=5.85(6) \lambda^2-10.64(8) \lambda+3.44(5) ,\qquad \qquad \label{semi_2_2}\\
c=-1.67(7)\lambda^2+3.34(10)\lambda-1.28(10), \qquad \qquad \label{semi_2_3}\\
d=0.67(3)\lambda-0.78(2) . \qquad \qquad \label{semi_2_4}
\end{eqnarray}

To extract the principal mass(es) from $E({\bf k})$ at a given value of $\lambda$,
we used the following procedure. At any $\lambda < \lambda_c$, we fitted the low-$k$ part of $E({\bf k})$ with Eqs. (\ref{BS11}) and/or (\ref{BS18}) and then used the relations (\ref{BS16}), (\ref{BS17}) and/or (\ref{BS19}),
(\ref{BS21})--(\ref{BS22}), respectively.
Equations (\ref{BS11}) and/or (\ref{BS18}) work for any $\lambda < \lambda_c$ since they properly
capture the Taylor expansion of $E({\bf k})$ at ${\bf k}=0$ up to the quartic terms inclusively.
At $\lambda > \lambda_c$ the range of applicability of such a protocol is finite, but
is still noticeably larger that the range of applicability of semi-analytic relations (\ref{semi_1_1})--(\ref{semi_1_4})
and/or  (\ref{semi_2_1})--(\ref{semi_2_4}); see Fig.~\ref{FIG7}. When fitting with ansatzes  (\ref{BS11}) and/or (\ref{BS18}) becomes poor, we fit with the generic finite-${\bf k}_0$ ansatz (\ref{BS4}) with the principal axes
(\ref{BS7}). The data produced with all the three protocols demonstrate perfect consistency; see Fig.~\ref{FIG7}.

The most significant quantitative difference between the adiabatic and intermediate regimes
is the values of the effective mass at strong coupling---for $\gamma = 3/8$ the heaviest
effective mass is within $20$\% of the bare mass value.

\subsection{Model B: Vibrational modes residing on lattice bonds}
\label{sec:sec42}

\begin{figure}[h]
\centering
\includegraphics[ trim=13.0cm 0cm 1.5cm 0cm, clip=true, width=0.6\textwidth]{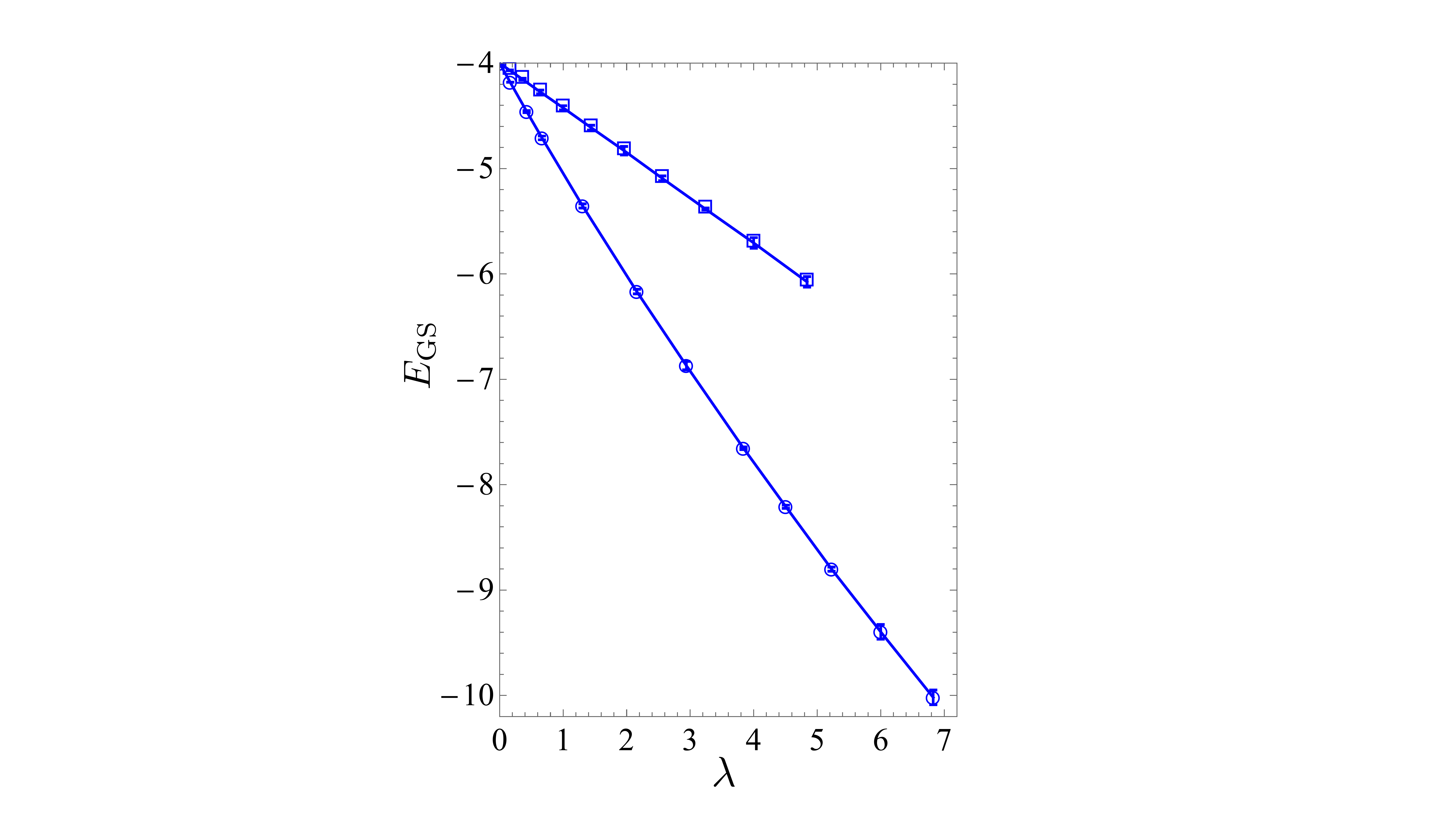}
\caption{(color online)
Ground state energy as a function of coupling strength $\lambda$ for model B in the
adiabatic, $\gamma=1/16$ (squares, upper curve), and intermediate $\gamma=3/8$ (circles, lower curve),
regimes.
}
\label{FIG1}
\end{figure}

In model B, particle hopping is modulated by the displacement of atoms located at lattice bonds.
We find that despite close similarities between models A and B in terms of physics involved,
fine details of the coupling vertex matter, both qualitatively and quantitatively.

\begin{figure}
\includegraphics [ trim=11cm 0cm 3cm 0cm, clip=true, width=0.6 \textwidth]{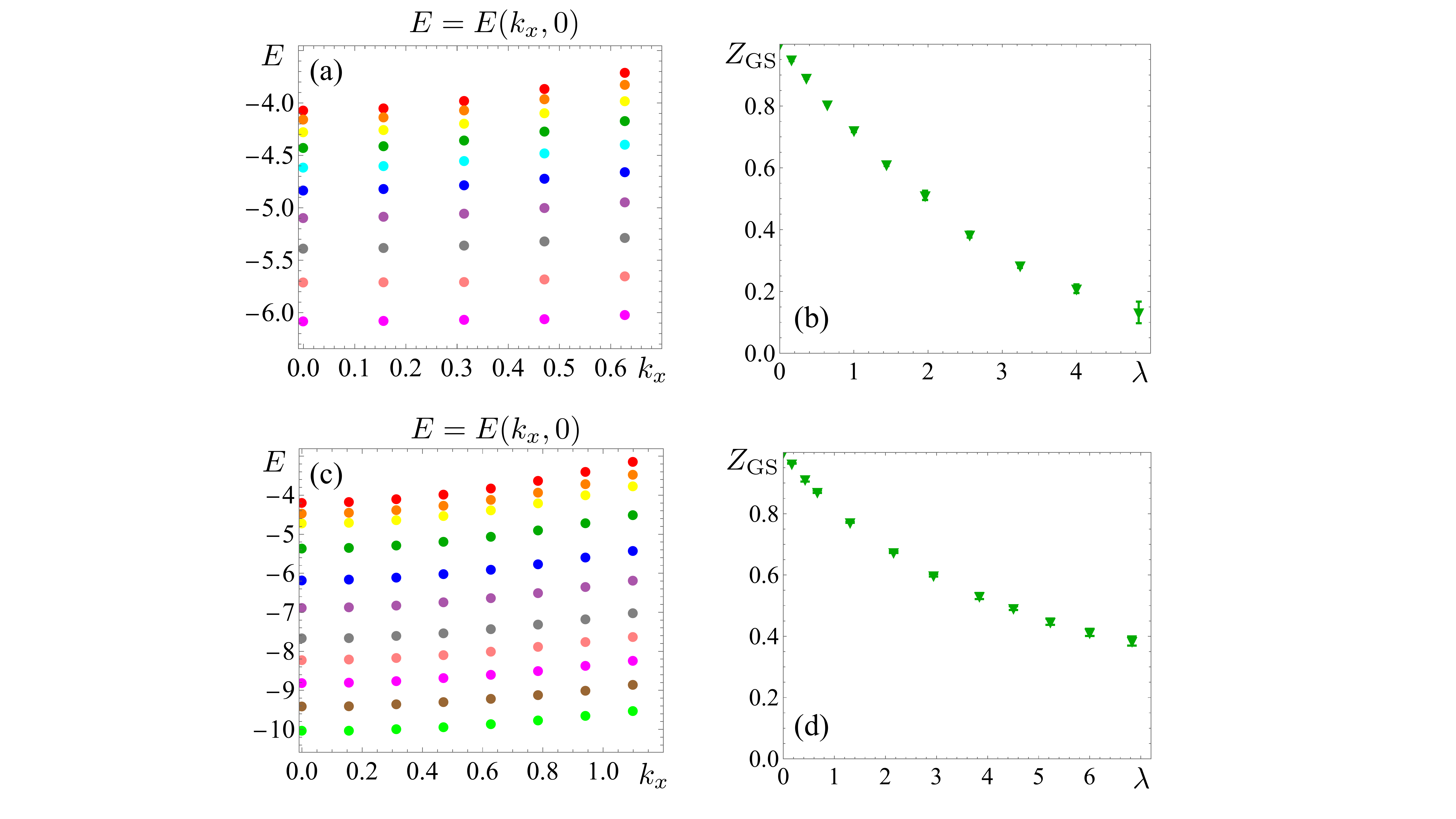}
\caption{(color online)
Polaron properties in the adiabatic, $\gamma=1/16$ (upper row), and intermediate $\gamma=3/8$ (lower row), regimes for model B. (a) Energy dispersion up to the decay threshold. Top to bottom: $\lambda=0.16$, 0.36, 0.64, 1.0, 1.44, 1.96, 2.56, 3.24, 4.0, and 4.84. (b) Ground state $Z_{\rm GS}$-factor. (c) Energy dispersion up to the decay threshold. Top to bottom: $\lambda=0.167$, 0.427, 0.667, 1.307, 2.16, 2.94, 3.84, 4.507, 5.227, 6.0, and 6.827. (d) Ground state $Z_{\rm GS}$-factor.
}
\label{FIG2}
\end{figure}

The quadratic-instability transition is absent in model B up to the largest coupling constant simulated,
and the ground state is always located at zero momentum. Since properties of the
sign-alternating expansion are more ``forgiving'' in this case, we were able to obtain data
for significantly larger ground state energy shifts, see Fig.~\ref{FIG1}. The
$E_{\rm GS} (\lambda)$ curves indicate that the ground state evolves smoothly with coupling. The energy dispersion
data presented in panels (a) and (c) in Fig.~\ref{FIG2} unambiguously confirm this conclusion by
demonstrating that the minimum at $\mathbf{k}=0$ is unique and its properties do not undergo rapid changes.
This is further evidenced by the $Z$-factor curves, see panels (b) and (d) in Fig.~\ref{FIG2}.

\begin{figure}
\centering
\includegraphics[ trim=14.0cm 0cm 1.5cm 0cm, clip=true, width=0.6\textwidth]{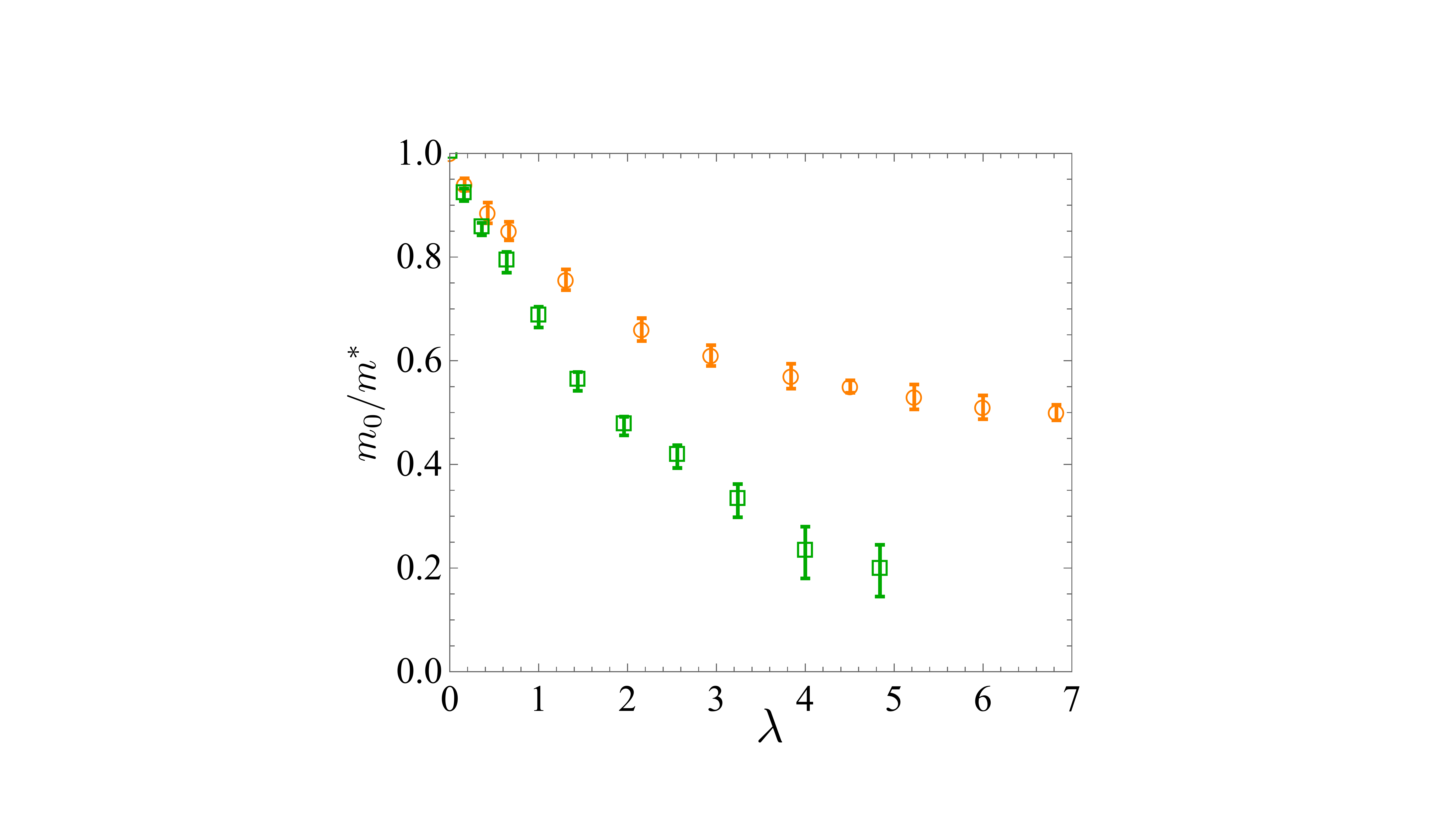}
\caption{(color online)
Effective mass as a function of coupling strength for model B in the
adiabatic, $\gamma =1/16$ (green squares), and intermediate, $\gamma =3/8$ (orange circls), regimes.
}
\label{FIG3}
\end{figure}

In the absence of quadratic instability, the effective mass renormalization in model B
remains modest all the way into the strong coupling regime, and, similarly to model A, appears
to level off as $\lambda$ is increased, see Fig.~\ref{FIG3}, in both adiabatic and intermediate regimes.
In Holstein model, for $\gamma=1/16$ the value of $m_0/m^*$ would be exponentially suppressed to near zero
for the same values of $E_{\rm GS}$.

Different properties of PSSH polarons in models A and B can be explained for
large $\omega_{\rm ph} /t$ as follows.
In model A, virtual excitation of the local phonon mode leads to the effective next-nearest-neighbor (n.n.n) hopping
amplitude with the negative sign \cite{Berciu:2010cq}. Indeed, consider a double-hopping event from site
$i$ to site $i+1$ with simultaneous excitation of the atomic vibration on site $i+1$
in the direction of hopping, let it be $\hat{x}$, and then to site $i+2$ (in the same direction)
with de-excitation of the same vibrational mode.
[There are no non-zero matrix elements to achieve the same goal for diagonal n.n.n. double-hopping amplitudes.]
Since matrix elements $\pm g$ for this process have opposite signs,
the second-order result for the effective transition amplitude is negative, $t_2 \approx g^2/\omega_{\rm ph} <0$.
[Recall that Hamiltonian matrix elements and hopping amplitudes have opposite sign, see Eq.~(\ref{t2})].
The dispersion relation based exclusively on $t$ and $t_2$ amplitudes
\begin{equation}
\begin{aligned}
E(\mathbf{k})-E_0 =&-& 2t [\cos (k_xa)+ \cos (k_ya)-2] \qquad
 \\
           \;      &-&2t_2[\cos(2k_x a)+\cos(2k_y a)-2] \, ,
\end{aligned}
\label{t2}
\end{equation}
has four symmetry-related  minima with $|k_{0x}| = |k_{0y}|$ for $t_2 < -t/4$.
In Eq.~(\ref{t2}) the new minima ``emerge'' from $\mathbf{k}=0$.

This argument does not work for model B, where de-excitation of the vibrational mode after the first
hopping event can only happen if the particle hops back to the same site; i.e., no large longer ranged
negative hopping amplitudes are generated.

\section{Conclusion}
\label{sec:sec6}

We investigated properties of polarons for two different two-dimensional PSSH Hamiltonians modeling
the electron-phonon interaction originating predominantly from hopping modulation by lattice vibrations.
Despite qualitative differences in some
ground state properties such as finite versus zero momentum, the two models share
an important feature: even in the adiabatic regime of small (compared to bandwidth) phonon frequencies,
the anisotropic effective mass renormalization is rather modest at strong coupling, in sharp contrast
with exponentially large effective mass observed in the Holstein model.
Light PSSH polarons, and subsequently bi-polarons (so far they were systematically studied
only in one dimension \cite{Sous:2018di}), offer a new perspective on the question of
bi-polaron mechanism of high-temperature superconductivity by eliminating the most serious
obstacle---exponentially large bi-polaron effective masses when they become energetically
stable. The other advantage comes from fundamentally non-local structure of
polaronic states in PSSH models, where electrons gain energy by hopping between the
lattice site. It is thus expected that PSSH bi-polarons will be less sensitive to
the repulsive local inter-electron interactions. The corresponding analysis is an important
direction for future work.

Since the superconducting transition temperature for bi-polarons increases with their density
one might assume that it is highest at half-filling. This is not necessarily the case because
of competing insulating crystalline states that emerge at commensurate filling factors
\cite{modelB}. The highest $T_c$ may correspond to a doped system.

\begin{acknowledgments}

NP and BS acknowledge support by the National Science Foundation
under Grant No. DMR-2032077. C. Zhang was supported by the
MURI Program ``New Quantum Phases of Matter'' from AFOSR. We thank the Supercomputing Center for Education $\&$ Research (OSCER) at the University of Oklahoma for providing us with their computational resources.
\end{acknowledgments}

\bibliography{PSSH}

\end{document}